\documentclass{appolb}
\usepackage[utf8x]{inputenc}

\usepackage{indentfirst}
\usepackage{array}
\usepackage{booktabs}
\usepackage{amsmath,amsfonts,amsthm,amssymb}
\usepackage[dvips]{graphics,graphicx}
\usepackage[usenames,dvipsnames]{color}
\definecolor{darkblue}{RGB}{0,0,196}
\definecolor{darkgreen}{RGB}{0,120,0}
\usepackage[colorlinks=true,linktocpage=true,linkcolor=darkblue,citecolor=red,urlcolor=darkblue]{hyperref}
\usepackage{cite}
\usepackage{widetext}
\usepackage{flushend}
\usepackage{cuted}
\usepackage{cancel}
\usepackage{bbold}
\usepackage{multirow}
\usepackage{longtable}
\usepackage{color}
\usepackage[normalem]{ulem}
\usepackage{hyperref}
\usepackage{bigints}
\usepackage{xparse}
\usepackage{physics}
\usepackage{verbatim}
\usepackage{minibox}
\usepackage{comment}
\usepackage{appendix}
\usepackage{slashed}
\usepackage{marginnote}
\usepackage{graphicx}
\usepackage[nice]{nicefrac}
\usepackage{amsmath}
\usepackage{textcomp}

\def\be{\begin{equation}}
\def\ee{\end{equation}}
\newcommand{\bel}[1]{\begin{eqnarray}\label{#1}}
\newcommand{\eel}{\end{eqnarray}}
\def\barr{\begin{array}}
\def\earr{\end{array}}
\def\beq{\begin{eqnarray}}
\def\eeq{\end{eqnarray}}
\def\bfig{\begin{figure}}
\def\efig{\end{figure}}

\newcommand{\nn}{\nonumber}

\def\S0iU{{\Sigma}^{0i}}

\def\n0{n_{0}}
\def\e0{\varepsilon_{0}}
\def\P0{P_{0}}

\def\be{\begin{equation}}
	\def\ee{\end{equation}}

\def\barr{\begin{array}}
	\def\earr{\end{array}}
\def\beq{\begin{eqnarray}}
	\def\eeq{\end{eqnarray}}
\def\bfig{\begin{figure}}
	\def\efig{\end{figure}}
\newcommand{\bea}{\begin{eqnarray}}
	\newcommand{\eea}{\end{eqnarray}}

\def\av{{\boldsymbol a}}
\def\bv{{\boldsymbol b}}

\def\be{\begin{equation}}
\def\ee{\end{equation}}
\def\ba{\begin{eqnarray}}
\def\ea{\end{eqnarray}}   

\def\av{{\boldsymbol a}}
\def\bv{{\boldsymbol b}}

\def\n0{n_{(0)}}
\def\e0{\varepsilon_{(0)}}
\def\P0{P_{(0)}}

\usepackage{stackengine,scalerel}
\newcommand\hstar[1]{\ThisStyle{\ensurestackMath{%
  \setbox0=\hbox{$\SavedStyle#1$}%
  \stackengine{0pt}{\copy0}{\kern.2\ht0\smash{\SavedStyle\star}}{O}{c}{F}{T}{S}}}}
\begin{document}

\title{Semi-classical kinetic theory for massive spin-half fermions with leading-order spin effects}
\author{Arpan Das
\address{Institute of Nuclear Physics Polish Academy of Sciences, PL-31-342 Krakow, Poland}\\
\smallskip
Wojciech Florkowski
\address{Institute of Theoretical Physics, Jagiellonian University, PL-30-348 Krakow, Poland}\\
\smallskip
Avdhesh Kumar
\address{Institute of Physics, Academia Sinica, Taipei, 11529, Taiwan}\\
\smallskip
Radoslaw Ryblewski
\address{Institute of Nuclear Physics Polish Academy of Sciences, PL-31-342 Krakow, Poland}\\
\smallskip
Rajeev Singh
\address{Institute of Nuclear Physics Polish Academy of Sciences, PL-31-342 Krakow, Poland}
\address{Center for Nuclear Theory, Department of Physics and Astronomy, Stony Brook University, Stony Brook, New York, 11794-3800, USA}}

\date{Received: date / Accepted: date}

\maketitle

\begin{abstract}
We consider the quantum kinetic-theory description for interacting massive spin-half fermions using the Wigner function formalism. We derive a general kinetic theory description assuming that the spin effects appear at the classical and quantum level. To track the effect of such different contributions we use the semi-classical expansion method to obtain the generalized dynamical equations including spin, analogous to classical Boltzmann equation. This approach can be used to obtain a collision kernel involving local as well as non-local collisions among the microscopic constituent of the system and eventually, a framework of spin hydrodynamics ensuring the conservation of the energy-momentum tensor and total angular momentum tensor.  
\end{abstract}
\section{Introduction}
\label{sec1:intro}
Relativistic fluid dynamics has been very successful in modeling the collective evolution of the strongly interacting matter produced in relativistic heavy-ion collision experiments~\cite{Florkowski:1321594,Gale:2013da,Jaiswal:2016hex,Romatschke:2017ejr,Florkowski:2017olj,Hidaka:2022dmn}.
Considering such success, an attempt has been made to incorporate the spin degrees of freedom within the hydrodynamic framework to explain the recent measurements of spin polarization of  particles emitted in these processes~\cite{STAR:2017ckg,Adam:2018ivw,STAR:2019erd,ALICE:2019aid,Acharya:2019ryw,Kornas:2019,STAR:2021beb,ALICE:2021pzu,STAR:2023eck}. Such a formalism of relativistic fluid dynamics with spin was first proposed in Ref.~\cite{Florkowski:2017ruc} and further developed in Refs.~\cite{Florkowski:2017dyn,Florkowski:2018ahw,Florkowski:2019qdp,Singh:2020rht,Bhadury:2020puc,Bhadury:2020cop,Singh:2021man,Florkowski:2021wvk,Ambrus:2022yzz,Singh:2022ltu, Daher:2022wzf,Biswas:2022bht,Xie:2023gbo,Wagner:2022amr,Weickgenannt:2023btk,Biswas:2023qsw,Bhadury:2023vjx,Jian-Hua:2023viv,Shi:2023sxh}; for reviews see Refs.~\cite{Florkowski:2018fap,Speranza:2020ilk,Becattini:2022zvf}.
Other similar approaches used concepts of thermodynamic equilibrium~\cite{Becattini:2009wh}, effective action~\cite{Montenegro:2018bcf,Montenegro:2020paq,Gallegos:2021bzp,Serenone:2021zef}, entropy current~\cite{Hattori:2019lfp,Fukushima:2020ucl,Li:2020eon,She:2021lhe,Daher:2022xon,Biswas:2023qsw}, statistical operator~\cite{Hu:2021lnx}, non-local collisions~\cite{Hidaka:2018ekt,Yang:2020hri,Wang:2020pej,Weickgenannt:2020aaf,Weickgenannt:2021cuo,Sheng:2021kfc,Weickgenannt:2022zxs,Hu:2021pwh,Hu:2022lpi,Weickgenannt:2022qvh}, chiral kinetic theory~\cite{Hidaka:2018ekt,Shi:2020htn,Huang:2023tyn}, and holographic duality~\cite{Gallegos:2020otk,Garbiso:2020puw,Gallegos:2022jow}.

The understanding of global polarization (along the direction of the total angular momentum) and local polarization (along the beam direction) phenomena has become a subject of very intense investigations recently~\cite{Gao:2020lxh,Becattini:2020ngo,lisa2021}.
While the theoretical approaches assuming spin-vorticity coupling~\cite{Becattini:2013fla} can explain the global polarization measurements~\cite{Becattini:2013fla,Becattini:2013vja,Becattini:2016gvu,Pang:2016igs,Xie:2017upb,Sun:2017xhx,Li:2017slc,Wei:2018zfb}, they fail to describe differential observables~\cite{Karpenko:2016jyx,Becattini:2017gcx,Xia:2018tes,Sun:2018bjl,Florkowski:2019voj}; though there are some recent advances in this respect \cite{Becattini:2021suc,Becattini:2021iol,Fu:2021pok,Florkowski:2021xvy,Sheng:2022ssd}.

On general thermodynamic grounds~\cite{Becattini:2018duy},
the spin polarization effects are expected to be quantified by an antisymmetric tensor~$\omega^{\mu \nu}$ conjugated to the generators of the Lorentz transformations, which in local thermal equilibrium
may be independent of the thermal vorticity~\cite{Florkowski:2017ruc,Florkowski:2017dyn,Florkowski:2018ahw}.
The spin polarization tensor $\omega^{\mu \nu}$ has been considered as a hydrodynamic variable that manifests the effects of spin at a macroscopic level. 
The hydrodynamic approach discussed in Refs.~\cite{Florkowski:2017ruc,Florkowski:2017dyn,Florkowski:2018ahw} implicitly assumes that the spin is a separately conserved quantity. 
However, in general, a microscopic collision process may also give rise to a transfer of angular momentum between the orbital and spin parts, keeping their sum conserved. An example of such a process is a non-local collision~\cite{Weickgenannt:2020aaf,Weickgenannt:2021cuo}.

In the current work, following Refs.~\cite{DeGroot:1980dk,Weickgenannt:2020aaf,Weickgenannt:2021cuo}, we extend the kinetic theory framework presented in Ref.~\cite{Florkowski:2018fap}. We use the Wigner function formalism to formulate a quantum kinetic theory for interacting spin-half Dirac particles~\cite{DeGroot:1980dk,Hakim:1379544}, and employ semi-classical approximation to derive transport equations of various components of the Wigner function~\cite{Elze:1986qd,Vasak:1987um,Elze:1989un,Zhuang:1995pd,Florkowski:1995ei}. However, considering a systematic $\hbar$ expansion of the Wigner function components, we assume that the spin polarization effects can be manifested at both leading and next-to-leading order, which goes beyond the situation discussed in Refs.~\cite{Weickgenannt:2020aaf,Weickgenannt:2021cuo}. 
The present approach can be used in future investigations to derive quantum kinetic equations using mapping of the Wigner function components to a classical distribution function with a phase-space extended to spin~\cite{Zamanian:2010zz,Ekman:2017kxi,Ekman:2019vrv}. Moreover, this approach can also be used to systematically include the effect of local and non-local collisions in the collision kernel~\cite{Weickgenannt:2020aaf,Weickgenannt:2021cuo}, which can be used to develop a spin hydrodynamic framework.

\smallskip

The structure of the paper is as follows: We begin with the description of the Wigner function formalism in Sec.~\ref{sec2:wigner function}.
Using the semi-classical expansion approach, in Sec.~\ref{sec2.2:wigner function}, we derive the transport equations for the components of the Wigner function, while in Sec.~\ref{sec:3:mass shell conditions} we obtain mass-shell conditions at zeroth and first-order in $\hbar$.
In Sec.~\ref{sec3:derivationkineticequations}, we derive kinetic equations for the scalar and axial-vector components, and in Sec.~\ref{sec:generalkinetic} we
formulate a general quantum kinetic equation.
We summarize our findings in Sec.~\ref{sec:summary} with possible future directions.
\smallskip

\textbf{Notations and Conventions:} In this work, we use the Cartesian coordinate system with $x^{\mu}\equiv (t,\boldsymbol{x})$ and the mostly-minus metric convention~\footnote{$g_{\mu\nu}$ = diag$(+1, -1, -1, -1)$.}.
The scalar product of two four-vectors $a$ and $b$ reads $a \cdot b = a^0 b^0 - \av \cdot \bv$, where the three-vectors are denoted by bold font. We work with the convention $\epsilon^{0123} =+1$ for the  Levi-Civita symbol $\epsilon^{\alpha\beta\gamma\delta}$. Also, for symmetrization and anti-symmetrization we use the notation $A_{\{\mu \nu\}} = A_{\mu\nu} + A_{\nu\mu}$ and $A_{[\mu \nu]} = A_{\mu\nu} - A_{\nu\mu}$, respectively.
We assume $c = k_B=1$ throughout while keeping $\hbar$ explicitly for our calculations.
The dual form of the tensor $A^{\mu\nu}$ is defined as
$\hstar{A}^{\mu\nu} = \frac{1}{2}\epsilon^{\mu\nu\alpha\beta}A_{\alpha\beta}$.
\section{Wigner function and its quantum kinetic equation}
\label{sec2:wigner function}
In the case of spin-half massive particles and the absence of gauge fields, the Wigner function can be expressed as follows~\cite{DeGroot:1980dk,Vasak:1987um}
\beq
W_{\alpha\beta}(x,k)=\int\frac{d^4y}{(2\pi\hbar)^4}~e^{-\frac{i}{\hbar}k\cdot y}\langle:\bar{\psi}_{\beta}(x_+)\psi_{\alpha}(x_-):\rangle,
 ~~\label{eq:WigFunc}
\eeq
where the Dirac field operator $\psi$ and its adjoint $\bar{\psi} \equiv \psi^{\dagger} \gamma^{0}$ are defined at two different points in spacetime $x_{\pm}\equiv x\pm y/2$ with $x$ and $y$
being the center and relative position, respectively. Here, the angle brackets indicate the ensemble average and the colon denotes normal ordering~\footnote{Note that alternative definitions of the Wigner function without normal ordering have also been considered in the literature, e.g. in Ref.~\cite{Gao:2019zhk}, where it has been argued that different definitions of the Wigner function may give rise to different results when the chiral anomaly is involved. However, we do not discuss such a situation here.}.

In the presence of interactions, the Dirac equation is expressed as~\cite{DeGroot:1980dk}
\beq
 \left( \slashed{p}-m\right)\!\psi(x)=\hbar\rho(x)\,,
\label{eq:DirEq}
\eeq
where $\rho(x)$ = $-(1/\hbar)\partial\mathcal{L}_I/\partial \bar{\psi}$, with $\mathcal{L}_{I}(x)$ denoting the interaction Lagrangian density, and $\slashed{p} =i\hbar \gamma^{\mu}\partial_{\mu}$. 

From the Lagrangian density
$\mathcal{L}(x)=\mathcal{L}_{D}(x)+\mathcal{L}_{I}(x)$
where  
\ba
\mathcal{L}_{D}(x)=\frac{1}{2} \bar{\psi}(x)\! \stackrel{\leftrightarrow}{\slashed{p}}\!\psi(x)-m \bar{\psi}(x) \psi(x)\,
\ea
is the Lagrangian density for the free Dirac field with mass $m$
and $\overleftrightarrow{\slashed{p}}  \equiv \overrightarrow{\slashed{p}} -\overleftarrow{\slashed{p}}$,
we can derive the following transport equation for the Wigner function \eqref{eq:WigFunc}
\cite{DeGroot:1980dk},
\beq
\left[\gamma \cdot k
+\frac{\slashed{p}}{2}-m\right]W(x,k)=\hbar~\mathcal{C}\left[W(x,k)\right].
 \label{equ3}
\eeq
Here, the collision term $\mathcal{C}\left[W(x,k)\right]$ is defined as~\cite{DeGroot:1980dk}
\beq
 \mathcal{C}\left[W(x,k)\right]\equiv \int \frac{d^4y}{(2\pi\hbar)^4}e^{-\frac{i}{\hbar}k \cdot y}\langle:\rho(x_-)\bar{\psi}(x_+):\rangle.
 \label{collisionterm}
\eeq
We would like to point out here that, in global equilibrium, the collision term \eqref{collisionterm} must vanish regardless of its form~\cite{DeGroot:1980dk,Hakim:1379544}. Here, we consider that the collision term describes the system away from equilibrium and gives rise to quantum corrections to the leading-order Wigner function, appearing at the $\hbar$ order or higher.

The Wigner function $W(x,k)$ is a matrix in the Dirac space, therefore we can express it in terms of the generators of the Clifford algebra as
\beq
W(x,k)&=&\frac{1}{4}\left[\boldsymbol{1} \, \mathcal{F}(x,k)+i \, \gamma^5 \, \mathcal{P}(x,k)+\gamma^{\mu} \, \mathcal{V}_{\mu}(x,k)\right.\nn\\
&+&\left.\gamma^5 \, \gamma^{\mu} \, \mathcal{A}_{\mu}(x,k)+\Sigma^{\mu\nu} \, \mathcal{S}_{\mu\nu}(x,k)\right],
 \label{equ5}
\eeq
with $\Sigma^{\mu\nu}\equiv (1/2) \sigma^{\mu\nu}\equiv (i/4)[\gamma^\mu,\gamma^\nu]$ being the Dirac spin operator.

Since the Wigner function is a complex matrix of order 4, it has 16 independent components: $\mathcal{F}(x,k)$, $\mathcal{P}(x,k)$, $\mathcal{V}_{\mu}(x,k)$, $\mathcal{A}_{\mu}(x,k)$,  and $\mathcal{S}_{\mu\nu}(x,k)$, which can be obtained by calculating the trace of $W(x,k)$ after multiplying first by the matrices:
${\Gamma_\mathcal{X}}\in\{\boldsymbol{1}, -i\gamma_5, \gamma^{\mu}, \gamma^{\mu}\gamma_5, 2\Sigma^{\mu\nu}\}$, where $\mathcal{X}\in \left\{ \mathcal{F}, \mathcal{P}, \mathcal{V}, \mathcal{A}, \mathcal{S}\right\}$, respectively. 

Under Lorentz transformations the expansion coefficients of the Wigner function, $\mathcal{F}$, $\mathcal{P}$, $\mathcal{V}_{\mu}$, $\mathcal{A}_{\mu}$  and $\mathcal{S}_{\mu\nu}$ transform as a scalar, pseudo-scalar, vector, axial-vector, and tensor, respectively~\cite{Vasak:1987um}.
The coefficients $\mathcal{F}$ and $\mathcal{P}$ have the interpretation of mass and pseudo-scalar condensate, respectively, whereas, $\mathcal{V}_{\mu}$ and $\mathcal{A}_{\mu}$ are known as the fermion number current density and the polarization density, respectively. Since $\mathcal{S}_{\mu\nu}$ is antisymmetric, it has six independent components having the physical interpretation of electric and magnetic dipole moments.

Using the representation of the Wigner function in terms of the generators of the Clifford algebra \eqref{equ5} in the kinetic equation \eqref{equ3} gives rise to kinetic equations for different coefficients of the Wigner function~\cite{Itzykson:1980rh}, which, after separating the real and imaginary parts, yields two sets of equations for $\mathcal{F},\mathcal{P},\mathcal{V}^{\mu}, \mathcal{A}^{\mu}$ and $\mathcal{S}^{\mu\nu}$, where the real parts are 
\ba
k^{\mu}\mathcal{V}_{\mu}-m \, \mathcal{F}&=&\hbar \, \mathcal{D}_{\mathcal{F}}\,,\label{equ6}\\
\frac{\hbar}{2} \, \partial^{\mu}\mathcal{A}_{\mu}+m \, \mathcal{P}&=&-\hbar \, \mathcal{D}_{\mathcal{P}}\,,\label{equ7}\\
k_{\mu}\mathcal{F}-\frac{\hbar}{2} \, \partial^{\nu}\mathcal{S}_{\nu\mu}-m \, \mathcal{V}_{\mu}&=&\hbar \, \mathcal{D}_{\mathcal{V},\mu}\,,\label{equ8}\\
 -\frac{\hbar}{2} \, \partial_{\mu}\mathcal{P}+k^{\beta}\hstar{\mathcal{S}}_{\mu\beta}
 +m \, \mathcal{A}_{\mu}&=&-\hbar \, \mathcal{D}_{\mathcal{A},\mu}\,,\label{equ9}\\
\frac{\hbar}{2} \, \partial_{[\mu}\mathcal{V}_{\nu]}-\epsilon_{\mu\nu\alpha\beta}k^{\alpha}\mathcal{A}^{\beta}-m \, \mathcal{S}_{\mu\nu}&=&\hbar \, \mathcal{D}_{\mathcal{S},{\mu\nu}}\label{equ10}\,,
\ea
while the imaginary parts are expressed as 
\ba
\hbar~\partial^{\mu}\mathcal{V}_{\mu}&=&2\hbar \, \mathcal{C}_{\mathcal{F}}\,,\label{equ11}\\
k^{\mu}\mathcal{A}_{\mu}&=&\hbar \, \mathcal{C}_{\mathcal{P}}\,,\label{equ12}\\
\frac{\hbar}{2}\partial_{\mu}\mathcal{F}+k^{\nu}\mathcal{S}_{\nu\mu}&=&\hbar \, \mathcal{C}_{\mathcal{V},\mu}\,,\label{equ13}\\   
k_{\mu}\mathcal{P}+\frac{\hbar}{2}\partial^{\beta}
\hstar{\mathcal{S}}_{\mu\beta}&=&-\hbar \,\mathcal{C}_{\mathcal{A},\mu}\,,\label{equ14}\\
k_{[\mu}\mathcal{V}_{\nu]}
+\frac{\hbar}{2}\epsilon_{\mu\nu\alpha\beta}\partial^{\alpha}\mathcal{A}^{\beta}&=&-\hbar \, \mathcal{C}_{\mathcal{S},\mu\nu}\,.\label{equ15}
\ea
In Eqs.~\eqref{equ6}-\eqref{equ15}, $\mathcal{D}_{\mathcal{X}}=\Re\text{Tr}\big[\Gamma_{\mathcal{X}}\mathcal{C}[W(x,k)]\big]$ and $\mathcal{C}_{\mathcal{X}}=\Im\text{Tr}\big[\Gamma_{\mathcal{X}}\mathcal{C}[W(x,k)]\big]$.

Note that since Eq.~\eqref{equ11} has $\hbar$ on both sides, one can argue that this equation can be considered at the leading order (zeroth order in $\hbar$), however, in this work we consider Eq.~\eqref{equ11} at the first order in $\hbar$~\cite{Vasak:1987um}, keeping $\hbar$ on both sides of the equation.
\section{Semi-classical expansion}
\label{sec2.2:wigner function}
In general, quantum kinetic equations~\eqref{equ6}--\eqref{equ15} are quite complicated because of the couplings between different components of the Wigner function. However, employing semi-classical expansion, we can decrease the complexity by breaking Eqs.~\eqref{equ6}--\eqref{equ15} into a number of independent equations.
The form of Eqs.~\eqref{equ6}--\eqref{equ15} indicate that we can search for the solutions for various components of the Wigner function in the form of a series expansion in $\hbar$,
${\mathcal{X}}=\sum_{n} \hbar^n{\mathcal{X}}^{(n)}$. Similarly, for the collision terms $\mathcal{C}_\mathcal{X}$ and $\mathcal{D}_\mathcal{X}$, we write
$\mathcal{C}_\mathcal{X}=\sum_n\hbar^{n}{\mathcal{C}^{(n)}_\mathcal{X}}$ and $\mathcal{D}_\mathcal{X}=\sum_n\hbar^{n}{\mathcal{D}^{(n)}_\mathcal{X}}$. Below we analyse Eqs.~\eqref{equ6}--\eqref{equ15} up to second-order in $\hbar$;
for the extension to the third-order,
see appendix \ref{Asec:2}.
\subsection{Zeroth order}
\label{subsec2.1:zeroth}
In the leading order, \textit{i.e.} the zeroth order in $\hbar$, the real parts give~\cite{Vasak:1987um}
\ba
k^{\mu}\mathcal{V}_{\mu}^{(0)}-m\mathcal{F}^{(0)}&=& 0\,,\label{equ33}\\
m\mathcal{P}^{(0)}&=&0\,,\label{equ34}\\
k_{\mu}\mathcal{F}^{(0)}-m\mathcal{V}_{\mu}^{(0)}&=&0\,,\label{equ35}\\
k^{\beta}\hstar{\mathcal{S}}_{\mu\beta}^{(0)}+m\mathcal{A}_{\mu}^{(0)}&=&0\,, \label{equ36}\\
\epsilon_{\mu\nu\alpha\beta}k^{\alpha}\mathcal{A}^{\beta(0)}+m\mathcal{S}_{\mu\nu}^{(0)}&=&0\,,  
\label{equ37}
\ea
while the imaginary parts yield~\cite{Vasak:1987um}
\ba
k^{\mu}\mathcal{A}_{\mu}^{(0)}&=&0\,,\label{equ38}\\
k^{\nu}\mathcal{S}_{\nu\mu}^{(0)}&=&0\,,\label{equ39}\\
k_{\mu}\mathcal{P}^{(0)}&=&0\,,\label{equ40}\\
k_{[\mu}^{}\mathcal{V}_{\nu]}^{(0)}
&=&0\,.\label{equ41}
\ea
From Eqs.~\eqref{equ33}--\eqref{equ41}, we conclude that $\mathcal{F}^{(0)}$ and $\mathcal{A}_{\mu}^{(0)}$ can be assumed as the basic independent coefficients in terms of which all other components of the Wigner function can be expressed, provided $\mathcal{A}_{\mu}^{(0)}$ satisfies orthogonality condition (\ref{equ38}).
\subsection{First order}
\label{subsec2.2:first}
In the first order in $\hbar$, real parts give
\ba
k^{\mu}\mathcal{V}_{\mu}^{(1)}-m\mathcal{F}^{(1)}&=&\mathcal{D}_{\mathcal{F}}^{(0)},\label{equ50}\\
\frac{1}{2}\partial^{\mu}\mathcal{A}_{\mu}^{(0)}+m\mathcal{P}^{(1)}&=&-\mathcal{D}_{\mathcal{P}}^{(0)},\label{equ51}\\
k_{\mu}\mathcal{F}^{(1)}-\frac{1}{2}\partial^{\nu}\mathcal{S}_{\nu\mu}^{(0)}-m\mathcal{V}_{\mu}^{(1)}&=& \mathcal{D}_{\mathcal{V},\mu}^{(0)},\label{equ52}\\
-\frac{1}{2}\partial_{\mu}\mathcal{P}^{(0)}+k^{\beta}\hstar{\mathcal{S}}_{\mu\beta}^{(1)}+m\mathcal{A}_{\mu}^{(1)}&=&-\mathcal{D}_{\mathcal{A},\mu}^{(0)},\label{equ53}\\
\frac{1}{2}\partial_{[\mu}\mathcal{V}_{\nu]}^{(0)}-\epsilon_{\mu\nu\alpha\beta}k^{\alpha}\mathcal{A}^{\beta(1)}-m\mathcal{S}_{\mu\nu}^{(1)}&=&\mathcal{D}_{\mathcal{S},{\mu\nu}}^{(0)}\label{equ54},
\ea
and the imaginary parts yield
\ba
\partial^{\mu}\mathcal{V}_{\mu}^{(0)}&=&2 \mathcal{C}_{\mathcal{F}}^{(0)},\label{equ55}\\
k^{\mu}\mathcal{A}_{\mu}^{(1)}&=&\mathcal{C}_{\mathcal{P}}^{(0)},\label{equ56}\\
\frac{1}{2}\partial_{\mu}\mathcal{F}^{(0)}+k^{\nu}\mathcal{S}_{\nu\mu}^{(1)}&=&\mathcal{C}_{\mathcal{V},\mu}^{(0)},\label{equ57}\\
k_{\mu}\mathcal{P}^{(1)}+\frac{1}{2}\partial^{\beta}
\hstar{\mathcal{S}}_{\mu\beta}^{(0)}&=&-\mathcal{C}_{\mathcal{A},\mu}^{(0)}, \label{equ58}\\
k_{[\mu}\mathcal{V}_{\nu]}^{(1)}
+\frac{1}{2}\epsilon_{\mu\nu\alpha\beta}\partial^{\alpha}\mathcal{A}^{\beta(0)}&=&- \mathcal{C}_{\mathcal{S},\mu\nu}^{(0)}.\label{equ59}
\ea
From Eq.~\eqref{equ56} we can immediately see that due to the presence of the collisions, the first-order axial-vector coefficient $\mathcal{A}^{(1)}$ is not orthogonal to $k$, cf. Eq.~\eqref{equ38}.
\subsection{Second order}
\label{subsec2.3:second}
We need to have second-order ($\mathcal{O}(\hbar^2)$) equations  of motion to derive the transport equations for the first-order ($\mathcal{O}(\hbar)$) coefficients of the Wigner function. Hence, in the second order, real parts give
\ba
k^{\mu}\mathcal{V}_{\mu}^{(2)}-m\mathcal{F}^{(2)}&=&\mathcal{D}_{\mathcal{F}}^{(1)},\label{equ68}\\
\frac{1}{2}\partial^{\mu}\mathcal{A}_{\mu}^{(1)}+m\mathcal{P}^{(2)}&=&-\mathcal{D}_{\mathcal{P}}^{(1)},\label{equ69}\\
k_{\mu}\mathcal{F}^{(2)}-\frac{1}{2}\partial^{\nu}\mathcal{S}_{\nu\mu}^{(1)}-m\mathcal{V}_{\mu}^{(2)}&=& \mathcal{D}_{\mathcal{V},\mu}^{(1)},\label{equ70}\\
-\frac{1}{2}\partial_{\mu}\mathcal{P}^{(1)}+k^{\beta}\hstar{\mathcal{S}}_{\mu\beta}^{(2)}+m\mathcal{A}_{\mu}^{(2)}&=&-\mathcal{D}_{\mathcal{A},\mu}^{(1)},\label{equ71}\\
\frac{1}{2}\partial_{[\mu}\mathcal{V}_{\nu]}^{(1)}-\epsilon_{\mu\nu\alpha\beta}k^{\alpha}\mathcal{A}^{\beta(2)}-m\mathcal{S}_{\mu\nu}^{(2)}&=&\mathcal{D}_{\mathcal{S},{\mu\nu}}^{(1)}\label{equ72},
\ea
while the imaginary parts yield
\ba
\partial^{\mu}\mathcal{V}_{\mu}^{(1)}&=&2 \mathcal{C}_{\mathcal{F}}^{(1)},\label{equ73}\\
k^{\mu}\mathcal{A}_{\mu}^{(2)}&=&\mathcal{C}_{\mathcal{P}}^{(1)},\label{equ74}\\
\frac{1}{2}\partial_{\mu}\mathcal{F}^{(1)}+k^{\nu}\mathcal{S}_{\nu\mu}^{(2)}&=&\mathcal{C}_{\mathcal{V},\mu}^{(1)}\,,\label{equ75}\\
k_{\mu}\mathcal{P}^{(2)}+\frac{1}{2}\partial^{\beta}
\hstar{\mathcal{S}}_{\mu\beta}^{(1)}&=&-\mathcal{C}_{\mathcal{A},\mu}^{(1)}\,, \label{equ76}\\
k_{[\mu}\mathcal{V}_{\nu]}^{(2)}
+\frac{1}{2}\epsilon_{\mu\nu\alpha\beta}\partial^{\alpha}\mathcal{A}^{\beta(1)}&=&- \mathcal{C}_{\mathcal{S},\mu\nu}^{(1)}\,.\label{equ77}
 \ea
\section{Mass-shell conditions}
\label{sec:3:mass shell conditions}
\subsection{Zeroth order}
\label{subsec:zerothorder}
From Eq.~\eqref{equ34} we trivially obtain the leading-order pseudo-scalar component,  $\mathcal{P}^{(0)}\!=\!0$, while from Eq.~\eqref{equ35} we find the expression for the leading-order vector coefficient in terms of the leading-order scalar coefficient as~\cite{Vasak:1987um,Florkowski:2018ahw,Gao:2020pfu}
\beq
     \mathcal{V}_{\mu}^{(0)}=\frac{k_{\mu}}{m}\mathcal{F}^{(0)}.
 \label{equ42.1}
 \eeq
Multiplying Eq.~\eqref{equ41} with $k^\mu$ and then using Eqs.~\eqref{equ33} and \eqref{equ35} we obtain constraint equation for the leading-order vector coefficient
\beq
 (k^2-m^2)\mathcal{V}_{\mu}^{(0)}=0.
 \label{equ49.1}
 \eeq
Inserting Eq.~\eqref{equ42.1} in Eq.~\eqref{equ33} we get analogous constraint equation for the leading-order scalar coefficient~\cite{Gao:2020pfu} 
\beq
    (k^2-m^2)\mathcal{F}^{(0)}=0.
    \label{equ46.1}
\eeq
From Eq.~\eqref{equ37} one can find the definition of the leading-order tensor coefficient and its dual in terms of leading-order axial-vector coefficient~\cite{Vasak:1987um,Florkowski:2018ahw}, respectively, as
\beq
\mathcal{S}_{\mu\nu}^{(0)} &=& -\frac{1}{m}\epsilon_{\mu\nu\alpha\beta}k^{\alpha}\mathcal{A}^{\beta}_{(0)},
    \label{equ51.1}\\
\hstar{\mathcal{S}}^{\mu\nu}_{(0)}
&=& \frac{1}{m} k^{[\mu}\mathcal{A}^{\nu]}_{(0)}.
\label{equ52.1}
\eeq
Using Eq.~\eqref{equ52.1} in Eq.~\eqref{equ36} and employing Eq.~\eqref{equ38} one can get the constraint equation for $\mathcal{A}_{\mu}^{(0)}$~\cite{Gao:2020pfu}
\ba
(k^2-m^2)\mathcal{A}_{\mu}^{(0)}&=&0.
\label{equ45}
\ea
One should note here, that due to our assumption that polarization effects can come from both the leading and first order in $\hbar$, $\mathcal{A}_{\mu}^{(0)}$ does not vanish which implies that the zeroth-order tensor coefficient $ \mathcal{S}_{\mu\nu}^{(0)}$ in Eq.~\eqref{equ51.1} does not vanish either. 
We would like to stress that in this respect our analysis goes beyond the approach of Refs.~\cite{Weickgenannt:2020aaf,Weickgenannt:2021cuo} and assumes that spin does not have to be only a dissipative effect, having both classical and quantum counterparts.

Similarly to Eq.~\eqref{equ51.1}, we can arrive at
the expression
\begin{eqnarray}
    \mathcal{A}^{\rho}_{(0)}=-\frac{k_{\lambda }}{m}{\hstar{\mathcal{S}}}^{\rho\lambda}_{(0)}=-\frac{1}{2 m} \epsilon ^{\rho \lambda \alpha \beta }k_{\lambda}\mathcal{S}_{\alpha\beta}^{(0)} \label{equ53.1}.
\end{eqnarray}
Putting Eq.~\eqref{equ53.1} in Eq.~\eqref{equ37}, and then using  Eq.~\eqref{equ39} we get the constraint equation for ${\mathcal{S}}_{\mu\nu}^{(0)}$
\beq
  (k^2-m^2){\mathcal{S}}_{\mu\nu}^{(0)}=0.\label{equ60.1}
 \eeq
Therefore, for a non-trivial solution to exist, all the leading-order coefficients have to satisfy the on-shell condition, \textit{i.e.}, $k^2=m^2$, where $k$ denotes the kinetic momentum.

Hence, in the leading order, we obtain Eqs.~\eqref{equ42.1}, \eqref{equ51.1}~and~\eqref{equ52.1}, in terms of independent quantities $\mathcal{F}^{(0)}$ and $\mathcal{A}_{\nu}^{(0)}$~\cite{Vasak:1987um,Florkowski:2018ahw}
where on-shell conditions for $\mathcal{F}^{(0)}$ and $\mathcal{A}_{\nu}^{(0)}$ (Eq.~\eqref{equ46.1}  and Eq.~\eqref{equ45}, respectively),  lead to~\cite{Gao:2020pfu}
\ba
\mathcal{F}^{(0)}=\delta(k^2-m^2){F}^{(0)},\quad
\mathcal{A}_{\mu}^{(0)}=\delta(k^2-m^2){A}_{\mu}^{(0)}\,,
~~~\label{equ:onshell}
\ea
with ${F}^{(0)}$ and ${A}_{\mu}^{(0)}$ being
arbitrary scalar and axial-vector functions, respectively, which are non-singular at $k^2=m^2$ and need to be determined by the kinetic equations.
One can easily verify at this point that Eqs.~\eqref{equ42.1}, \eqref{equ51.1}~and~\eqref{equ52.1} satisfy Eqs.~\eqref{equ33}--\eqref{equ41} if the axial-vector component of zeroth order fulfills the orthogonality condition~\eqref{equ38}.
\subsection{First order}
\label{subsec:firstorder}
From Eqs.~\eqref{equ51}, \eqref{equ52} and \eqref{equ54} one obtains the first-order contributions to the pseudo-scalar, vector and tensor components of the Wigner function, respectively, as
\ba
 \mathcal{P}^{(1)}&=&-\frac{1}{2m}\left[\partial^{\mu}\mathcal{A}_{\mu}^{(0)}+2\mathcal{D}_{\mathcal{P}}^{(0)}\right]\label{equ60},\\
 \mathcal{V}_{\mu}^{(1)}&=&\frac{1}{m}\left[k_{\mu}\mathcal{F}^{(1)}-\frac{1}{2}\partial^{\nu}\mathcal{S}_{\nu\mu}^{(0)}-\mathcal{D}^{(0)}_{\mathcal{V},\mu}\right]\label{equ61},\\
 \mathcal{S}_{\mu\nu}^{(1)}&=&\frac{1}{2m}\left[\partial_{[\mu}\mathcal{V}_{\nu]}^{(0)}-2\epsilon_{\mu\nu\alpha\beta}k^{\alpha}\mathcal{A}^{\beta}_{(1)}-2\mathcal{D}_{\mathcal{S},{\mu\nu}}^{(0)}\right]\label{equ62},
\ea
where the dual form of $\mathcal{S}_{\mu\nu}^{(1)}$ is obtained by contracting with Levi-Civita tensor,
\beq
    \hstar{\mathcal{S}}_{\mu\beta}^{(1)}&=&\frac{1}{m}\Big[\frac{1}{4}\epsilon_{\mu\beta \sigma \rho} \partial^{[\sigma}\mathcal{V}^{\rho]}_{(0)} + k_{[\mu}\mathcal{A}_{\beta]}^{(1)}-\frac{1}{2}\epsilon_{\mu\beta \sigma \rho }\mathcal{D}_{\mathcal{S}(0)}^{\sigma\rho}\Big].\nn\\
    \label{equ63}
\eeq
Using Eqs.~\eqref{equ34} and \eqref{equ63} in Eq.~\eqref{equ53} and subsequently using Eqs.~\eqref{equ56} and \eqref{equ42.1} we get constraint condition for first-order axial-vector coefficient as
\beq
(k^2-m^2)\mathcal{A}_{\mu}^{(1)}=k_{\mu}\mathcal{C}_{\mathcal{P}}^{(0)}-\frac{1}{2}\epsilon_{\mu\beta \sigma\rho}k^{\beta}\mathcal{D}_{\mathcal{S}(0)}^{\sigma\rho}+m\mathcal{D}_{\mathcal{A},\mu}^{(0)}.\nn\\
\label{equ64}
\eeq
To obtain the constraint equation satisfied by the first-order scalar coefficient
we contract Eq.~\eqref{equ52} with $k^{\mu}$ and use Eqs.~\eqref{equ50} and \eqref{equ51.1}, getting
\beq
(k^2-m^2)\mathcal{F}^{(1)}=k^{\mu}\mathcal{D}_{\mathcal{V},\mu}^{(0)}+m\mathcal{D}_{\mathcal{F}}^{(0)}.
\label{equ65}
\eeq
Multiplying Eq.~\eqref{equ59} with $k_{\mu}$ and then using Eq.~\eqref{equ50}, as well as Eqs.~\eqref{equ37} and  \eqref{equ52} yields the constraint equation for first-order vector coefficient 
    \beq
    & \big(k^2-m^2\big)\mathcal{V}^{\rho}_{(1)}=m\mathcal{D}^{\rho}_{\mathcal{V}(0)}+k^{\rho}\mathcal{D}_{\mathcal{F}(0)}^{}-k_{\lambda}\mathcal{C}^{\lambda\rho}_{\mathcal{S}(0)}.
    \label{equ103ver2}
\eeq
On the other hand, multiplying Eq.~\eqref{equ58} by $k_{\rho}$ 
we get
\beq
   k^2\mathcal{P}^{(1)}+\frac{1}{2} k_{\rho}\partial_{\lambda}\hstar{\mathcal{S}}^{\rho\lambda}_{(0)}=-k_{\rho}\mathcal{C}^{\rho}_{\mathcal{A}(0)}.
    \label{equ104ver1}
\eeq
Subtracting Eq.~\eqref{equ51} multiplied by $m$
from Eq.~\eqref{equ104ver1}, and using in the resulting formula Eq.~\eqref{equ36}, we get the constraint equation for first-order pseudo-scalar coefficient as
\beq
(k^2-m^2)\mathcal{P}^{(1)}  =  -k_{\rho}\mathcal{C}^{\rho}_{\mathcal{A}(0)}+m\mathcal{D}_{\mathcal{P}}^{(0)}.
\label{equ106ver1}
\eeq
Combining Eqs.~\eqref{equ34}, \eqref{equ53} and \eqref{equ54}, after some straightforward algebraic manipulations, gives us the following equation
\beq
 m \partial^{[\rho}\mathcal{V}^{\lambda]}_{(0)}
 -\epsilon^{\alpha\rho\lambda\sigma}\epsilon_{\alpha\gamma\delta\beta}k_{\sigma}k^{\beta}\mathcal{S}^{\gamma\delta}_{(1)}
    + 2 \epsilon^{\rho\lambda\sigma\alpha}k_{\sigma}\mathcal{D}^{(0)}_{\mathcal{A},\alpha}-2 m^2 \mathcal{S}^{\rho\lambda}_{(1)} = 2 m \mathcal{D}^{\rho\lambda}_{\mathcal{S}(0)}.
    \label{equ108ver1}
\eeq
Contracting Levi-Civita tensors and using Eq.~\eqref{equ57} in   Eq.~\eqref{equ108ver1}, and subsequently using Eq.~\eqref{equ35}, one arrives at
the constraint equation for first-order tensor coefficient $\mathcal{S}^{\rho\lambda}_{(1)}$, 
\beq
(k^2-m^2)\mathcal{S}^{\rho\lambda}_{(1)} = k_{}^{[\rho}\mathcal{C}_{\mathcal{V}(0)}^{\lambda]}+m\mathcal{D}^{\rho\lambda}_{\mathcal{S}(0)}-\epsilon^{\rho\lambda\sigma\alpha}k_{\sigma}\mathcal{D}^{(0)}_{\mathcal{A},\alpha}.
~~~~~\label{equ112ver2}
\eeq
Eqs.~\eqref{equ64}--\eqref{equ103ver2}, \eqref{equ106ver1} and \eqref{equ112ver2} are the mass-shell conditions for all the first-order coefficients of the Wigner function.
From these equations, one can observe that in the collisionless limit, all first-order coefficients remain on-shell. Furthermore, unlike in Refs.~\cite{Weickgenannt:2020aaf,Weickgenannt:2021cuo}, the above equations assume the most general structure of the interactions by considering all the collision terms to be nonvanishing, \emph{i.e.} $\mathcal{D}_{\mathcal{X}}\neq 0$ and $\mathcal{C}_{\mathcal{X}}\neq 0$.
\section{Kinetic equations for scalar and axial-vector components}
\label{sec3:derivationkineticequations}
Combining  Eqs.~\eqref{equ55} and \eqref{equ42.1} we find the kinetic equation to be satisfied by the leading-order scalar coefficient
\beq
k^{\mu}\partial_{\mu}\mathcal{F}^{(0)}=2m\,\mathcal{C}_{\mathcal{F}}^{(0)},
    \label{equ66}
\eeq
while plugging Eq.~\eqref{equ61} in Eq.~\eqref{equ73} we obtain the kinetic equation for the first-order scalar coefficient
\beq
k^{\mu}\partial_{\mu}\mathcal{F}^{(1)}=2m\,\mathcal{C}_{\mathcal{F}}^{(1)}+\partial^{\mu}\mathcal{D}_{\mathcal{V},\mu}^{(0)}.\label{equ66.1}
\eeq
Using Eq.~\eqref{equ51} together with Eq.~\eqref{equ52.1} in Eq.~\eqref{equ58}, we get the kinetic equation to be satisfied by leading-order axial-vector coefficient 
\begin{eqnarray}
k^{\beta}\partial_{\beta}\mathcal{A}_{\mu}^{(0)}&=& 2m\,\mathcal{C}_{\mathcal{A},\mu}^{(0)}-2k_{\mu}\mathcal{D}_{\mathcal{P}}^{(0)}.
\label{equ67} 
\end{eqnarray}
Finally, using Eq.~\eqref{equ63} and  Eq.~\eqref{equ78} in Eq.~\eqref{equ76} we get the kinetic equation to be satisfied by the first-order axial-vector coefficient 
\beq
    k^{\beta}\partial_{\beta}\mathcal{A}_{\mu}^{(1)}=2m\,\mathcal{C}_{\mathcal{A},\mu}^{(1)}-2k_{\mu}\mathcal{D}_{\mathcal{P}}^{(1)}-\frac{1}{2}\epsilon_{\mu\beta \gamma\delta}\partial^{\beta}\mathcal{D}_{\mathcal{S}(0)}^{\gamma\delta}.
~~~\label{equ86}
\eeq
So far, we have discussed mass-shell conditions and kinetic equations satisfied by different components of the  Wigner function without imposing any special assumptions. At this point, it is useful to highlight the important differences between the present work and Refs.~\cite{Weickgenannt:2020aaf,Weickgenannt:2021cuo}.
The crucial assumption considered in Refs.~\cite{Weickgenannt:2020aaf,Weickgenannt:2021cuo} is that the spin effects appear at first order in $\hbar$ in the semi-classical expansion of the Wigner function.
Such an assumption is physically motivated when one considers relatively small spin polarization effects arising through scatterings in a vortical medium.
Within the quantum kinetic theory approach, the polarization effects appear through the axial-vector component
$\mathcal{A}^{\mu}$ \cite{Weickgenannt:2019dks,Weickgenannt:2020aaf,Weickgenannt:2021cuo}.
The assumption that the spin polarization effect is at least of order $\hbar$ implies that $\mathcal{A}^{(0)}_{\mu}$ can be considered to be zero, and consequently, the tensor component of the Wigner function $\mathcal{S}^{(0)}_{\mu \nu}=0$ (see Eq.~\eqref{equ51.1}).
Moreover, the leading-order pseudo-scalar component $\mathcal{P}^{(0)}$ is always vanishing~\eqref{equ34}.
Furthermore, for the consistency of the framework, it has been argued that the leading-order collision terms involving the pseudo-scalar, axial-vector, and tensor components must vanish, \textit{i.e.}, $\mathcal{C}_{\mathcal{P}}^{(0)}=0$; $\mathcal{C}_{\mathcal{A}}^{\mu(0)}=0$;
$\mathcal{C}_{\mathcal{S}}^{\mu\nu(0)}=0$;
$\mathcal{D}_{\mathcal{P}}^{(0)}=0$;
$\mathcal{D}_{\mathcal{A}}^{\mu(0)}=0$;
$\mathcal{D}_{\mathcal{S}}^{\mu\nu(0)}=0$~\cite{Weickgenannt:2020aaf,Weickgenannt:2021cuo}.
Such constraints on the collision terms also affect the on-shell conditions for various components of the Wigner function.
It can be shown in this case that  $\mathcal{A}_{\mu}^{(1)}$ not only is on-shell, see Eq.~\eqref{equ64}, but also remains orthogonal to momentum, $k^{\mu}\mathcal{A}_{\mu}^{}=\mathcal{O}(\hbar^2)$~\cite{Weickgenannt:2020aaf,Weickgenannt:2021cuo}. Similarly, due to the assumptions that $\mathcal{A}_{\mu}^{(0)}$ and the collision term $\mathcal{D}_{\mathcal{P}}^{(0)}$ are vanishing, the pseudo-scalar component $\mathcal{P}$ is at least of the order of $\hbar^2$, see Eq.~\eqref{equ60}.

The novelty of the current study in comparison to Refs.~\cite{Weickgenannt:2020aaf,Weickgenannt:2021cuo}, is that we have assumed that the polarization effects can also appear at the leading-order of the semi-classical approximation, \textit{i.e.} $\mathcal{A}^{(0)}_{\mu}\neq 0$.
The physical motivation behind our assumption is that the polarization effect can be manifested even at the classical level, and thus polarization can be generated in the presence as well as in the absence of specific collision processes in a vortical fluid.
We also assume that all the collision terms are, in general, non-vanishing, hence, in contrast to
Refs.~\cite{Weickgenannt:2020aaf,Weickgenannt:2021cuo}, we consider $\mathcal{P}^{(1)}\neq 0$,
$(k^2-m^2)\mathcal{A}_{\mu}^{(1)}\neq 0$,
and $k^{\mu}\mathcal{A}_{\mu}=\mathcal{O}(\hbar)$. Such a different treatment of the axial-vector component of the Wigner function and various collision terms may give rise to significantly different on-shell conditions and kinetic equations for various components of the Wigner function.
\section{General Kinetic equation and its classical counterpart}
\label{sec:generalkinetic}
In this section we combine the zeroth and first-order kinetic equations for $\mathcal{F}$ and $\mathcal{A}_{\mu}$.
This may be used to develop a hydrodynamic framework where spin effects arise at both $\hbar^0$ and $\hbar^1$ orders.

Combining Eqs.~\eqref{equ66} and \eqref{equ66.1} we get the kinetic equation for the scalar coefficient as
\begin{eqnarray}
    k^{\mu}\partial_{\mu}\tilde{\mathcal{F}}&=&2m \,\tilde{\mathcal{C}}_{\mathcal{F}}\,,
    \label{equ:gke2}
\end{eqnarray}
where 
\begin{eqnarray}
   \tilde{\mathcal{F}}&=&\mathcal{F}^{(0)}+\hbar\,\mathcal{F}^{(1)},
  \nn\\
\tilde{\mathcal{C}}_{\mathcal{F}}&=&\mathcal{C}_{\mathcal{F}}^{(0)}+\hbar\left(\mathcal{C}_{\mathcal{F}}^{(1)}+\frac{1}{2m}\partial^{\mu}\mathcal{D}_{\mathcal{V},\mu}^{(0)}\right).
\label{equ:gke4}
\end{eqnarray}
Similarly, from Eqs.~\eqref{equ67} and \eqref{equ86}, we obtain the following kinetic equation for axial-vector component
\begin{eqnarray}
    k^{\beta}\partial_{\beta}\tilde{\mathcal{A}}_{\mu}&=&2m~\tilde{\mathcal{C}}_{\mathcal{A},\mu}\,, \label{equ:gke1.2}
\end{eqnarray}
where
\begin{eqnarray}
  \tilde{\mathcal{A}}_{\mu}&=&\mathcal{A}_{\mu}^{(0)}+\hbar\,\mathcal{A}_{\mu}^{(1)}, \label{equ:gke1.3}\\
  \tilde{\mathcal{C}}_{\mathcal{A},\mu} &=&\mathcal{C}_{\mathcal{A},\mu}^{(0)}+\hbar \,  \mathcal{C}_{\mathcal{A},\mu}^{(1)}-\frac{k_{\mu}}{m}\left(\mathcal{D}_{\mathcal{P}}^{(0)}+\hbar \mathcal{D}_{\mathcal{P}}^{(1)}\right)\nn\\
  &-& \frac{\hbar}{4m}\epsilon_{\mu\beta \gamma\delta}\partial^{\beta}\mathcal{D}_{\mathcal{S}(0)}^{\gamma\delta}.
  \label{equ:gke1.4}
\end{eqnarray}
It is useful to introduce spin as an additional phase-space variable in the distribution function as~\cite{Zamanian:2010zz,Ekman:2017kxi,Florkowski:2018fap,Ekman:2019vrv,Weickgenannt:2020aaf}
\begin{eqnarray}
  \mathfrak{f}(x,k,\mathfrak{s})=\frac{1}{2}\left(\tilde{\mathcal{F}}(x,k)-\mathfrak{s}\cdot\tilde{\mathcal{A}}(x,k)\right) ,\label{equ:fspin}
\end{eqnarray}
where  $\mathfrak{s}^{\alpha}$ is the spin four-vector. Using Eq.~\eqref{equ:fspin} one can also obtain $\tilde{\mathcal{F}}(x,k)$ and $\tilde{\mathcal{A}}(x,k)$, 
\begin{align}
    \int \mathrm{dS}(k) \, \mathfrak{f}(x,k,\mathfrak{s}) &= \tilde{\mathcal{F}}(x,k)\,,
    \label{equ79ver2}\\
    \int \mathrm{dS}(k) \, \mathfrak{s}^{\mu}~\mathfrak{f}(x,k,\mathfrak{s}) &= \tilde{\mathcal{A}}^{\mu}(x,k)\,,
    \label{equ80ver2}
\end{align}
where the spin measure~\cite{Weickgenannt:2020aaf,Weickgenannt:2021cuo}
\begin{align}
    \int \mathrm{dS}(k) \equiv \frac{1}{\pi}\sqrt{\frac{k^2}{3}} \int d^4\mathfrak{s} \, \delta(\mathfrak{s}\cdot \mathfrak{s}+3) \, \delta(k\cdot \mathfrak{s})\,,
    \label{eq:spin_measure}
\end{align}
satisfies the following identities~\cite{Weickgenannt:2020aaf,Weickgenannt:2021cuo}
\begin{align}
\int \mathrm{dS}(k) &= 2\,,\nn\\
\int \mathrm{dS}(k)    \, \mathfrak{s}^{\mu} &= 0\,,\nn\\
\int \mathrm{dS}(k) \, \mathfrak{s}^{\mu}\mathfrak{s}^{\nu} &=
-2\bigg(g^{\mu\nu}-\frac{k^{\mu}k^{\nu}}{k^2}\bigg)\,.
\end{align}
The relation between $\tilde{\mathcal{F}}(x,k)$ and $\mathfrak{f}(x,k,\mathfrak{s})$
\eqref{equ79ver2} can be easily established using the above identities.
But it is rather non-trivial to express $\tilde{\mathcal{A}}^{\mu}(x,k)$ in terms of $\mathfrak{f}(x,k,\mathfrak{s})$.
Using Eq.~\eqref{equ:fspin} on the left-hand side of Eq.~\eqref{equ80ver2} it can be shown that, 
\begin{align}
    \int \mathrm{dS}(k) \, \mathfrak{s}^{\mu} \, \mathfrak{f}(x,k,\mathfrak{s}) = \tilde{\mathcal{A}}^{\mu}(x,k)-\frac{k^{\mu}}{k^2}\left(k\cdot\tilde{\mathcal{A}}\right). 
    \label{amuinversion}
\end{align}
Equation \eqref{equ80ver2} is valid only when $k \cdot \tilde{\mathcal{A}}=0$.
From Eqs.~\eqref{equ38} and \eqref{equ56} one can observe that $k \cdot \tilde{\mathcal{A}}=\hbar \, \mathcal{C}_{\mathcal{P}}^{(0)}$.
Thus the collision term $\mathcal{C}_{\mathcal{P}}^{(0)}$ must vanish to obtain Eq.~\eqref{equ80ver2}. In Refs.~\cite{Weickgenannt:2020aaf,Weickgenannt:2021cuo} it was considered that $\mathcal{C}_{\mathcal{P}}^{(0)}=0$ since the spin polarization effects are assumed to be at least of order $\hbar$.
In the present investigation one can still consider that $\mathcal{C}_{\mathcal{P}}^{(0)}$ vanishes.
From Eqs.~\eqref{equ9} and \eqref{equ12} one obtains
\begin{align}
  k^{\mu}\partial_{\mu}\mathcal{P}=2m~\mathcal{C}_{\mathcal{P}}+2k^{\mu} \mathcal{D}_{\mathcal{A},\mu}\,.
 \end{align}
Using the semi-classical expansion of $\mathcal{P}$, $\mathcal{C}_{\mathcal{P}}$ and $\mathcal{D}_{\mathcal{A},\mu}$ with the condition $\mathcal{P}^{(0)}=0$ the above equation gives
\begin{align}
m \, \mathcal{C}_{\mathcal{P}}^{(0)} + k^{\mu} \mathcal{D}_{\mathcal{A},\mu}^{(0)}=0.
 \label{equ87ver2}
\end{align}
If we consider that the collision term $\mathcal{D}_{\mathcal{A},\mu}^{(0)}$ is such that it satisfies $k^{\mu} \mathcal{D}_{\mathcal{A},\mu}^{(0)}=0$, then Eq.~\eqref{equ87ver2} gives $\mathcal{C}_{\mathcal{P}}^{(0)} = 0$ \footnote{Note that  the collision term $\mathcal{D}_{\mathcal{A},\mu}^{(0)}$ transforms as an axial-vector under the Lorentz transformation. Microscopically such axial vector nature of $\mathcal{D}_{\mathcal{A},\mu}^{(0)}$ may originate from spin ($\mathfrak{s}^{\mu}$) which transforms like a axial vector. Therefore, one may express $\mathcal{D}_{\mathcal{A},\mu}^{(0)}$ in terms of the spin four-vector $\mathfrak{s}^{\mu}$. Since $k \cdot \mathfrak{s} = 0$ due to the Dirac delta function $\delta(k\cdot\mathfrak{s})$ in the expression of generalized spin measure \eqref{eq:spin_measure}, one may expect that $k^{\mu} \mathcal{D}_{\mathcal{A},\mu}^{(0)}=0$ for a generic collision term.}. In that case, Eqs.~\eqref{equ79ver2} and \eqref{equ80ver2} provide a well defined prescription to obtain $\tilde{\mathcal{F}}(x,k)$ and $\tilde{\mathcal{A}}^{\mu}(x,k)$ in terms of the distribution function $\mathfrak{f}(x,k,\mathfrak{s})$ \footnote{One may also define, alternatively, the distribution function \eqref{equ:fspin} as $\mathfrak{f}(x,k,\mathfrak{s})=\frac{1}{2}\left(\tilde{\mathcal{F}}(x,k)-\mathfrak{s}\cdot\tilde{\mathcal{A}}_{\perp}(x,k)\right)$ where $\tilde{\mathcal{A}}_{\perp}(x,k)$ is the orthogonal component of $\tilde{\mathcal{A}}_{}(x,k)$ with respect to $k^{\mu}$, i.e. $\tilde{\mathcal{A}}^{\mu}_{\perp}(x,k)= \tilde{\mathcal{A}}^{\mu}(x,k)-\frac{k^{\mu}}{k^2}\left(k\cdot\tilde{\mathcal{A}}\right)$. 
In that case, Eqs.~\eqref{equ79ver2} and \eqref{equ80ver2} are trivially satisfied without any constraint on the collision terms.}. The inversion relations from $\tilde{\mathcal{F}}(x,k)$ and $\tilde{\mathcal{A}}^{\mu}(x,k)$ to $\mathfrak{f}(x,k,\mathfrak{s})$ are crucial because, in the quantum kinetic theory approach, energy-momentum and the spin tensors are expressed in terms of various independent components of the Wigner function.
Therefore, with the help of inversion relations 
\eqref{equ79ver2} and \eqref{equ80ver2}, those macroscopic currents can be written in terms of $\mathfrak{f}(x,k,\mathfrak{s})$ \cite{Speranza:2020ilk,Florkowski:2017ruc,Florkowski:2017dyn,Florkowski:2018ahw}.
From Eqs.~\eqref{equ:gke2}, \eqref{equ:gke1.2} and \eqref{equ:fspin}, we get the general Boltzmann equation to be solved~\footnote{Note that Boltzmann equation (\ref{eq:kineticspin}) derived here, assumes that spin effects are at least of leading and first-order in $\hbar$ which is not the case derived in Refs.~\cite{Weickgenannt:2020aaf,Weickgenannt:2021cuo}.}
\begin{eqnarray}
k^{\mu}\partial _{\mu }\mathfrak{f}(x,k,\mathfrak{s})=m\, \mathfrak{C}(\mathfrak{f}), \label{eq:kineticspin}
\end{eqnarray}
where $\mathfrak{C}(\mathfrak{f})\!=\! \tilde{\mathcal{C}}_{\mathcal{F}}-\mathfrak{s}\cdot \tilde{\mathcal{C}}_{\mathcal{A}}$ is the collision term that explicitly takes into account the effect of spin.
Within the quasiparticle approximation, one may choose a generalized representation of the distribution function  $\mathfrak{f}(x,k,\mathfrak{s})=m\delta(k^2-M^2)f(x,k,\mathfrak{s})$~\cite{Weickgenannt:2020aaf,Weickgenannt:2021cuo}.
Here, the on-shell singularity for the quasiparticle has been expressed as $\delta(k^2-M^2)$ and the function $f(x,k,\mathfrak{s})$ is a function without singularity.
Furthermore, $M$ denotes quasiparticle mass, which includes quantum corrections to the bare mass $m$. These quantum corrections (off-shell corrections) can primarily originate from the spin dependence. Note that, these off-shell contributions can appear on both sides of the above equation. If the off-shell terms from both sides of Eq.~\eqref{eq:kineticspin} cancel then one will be left with the on-shell Boltzmann equation involving the distribution function $f(x,k,\mathfrak{s})$~\cite{Weickgenannt:2020aaf,Weickgenannt:2021cuo}. To check this cancellation of off-shells correction one has to calculate the generalized collision term $\mathfrak{C}(\mathfrak{f})$ involving spin-dependent distribution function in the presence of local and non-local collisions. Explicit expressions of such spin-dependent collision kernels have been put forward in Refs.~\cite{Weickgenannt:2020aaf,Weickgenannt:2021cuo} where one only considers the spin contribution at the order of $\hbar$.
The explicit expression of $\mathfrak{C}(\mathfrak{f})$ in Eq.~\eqref{eq:kineticspin}
that incorporate spin effects in both the leading and the first order in $\hbar$ 
including local and non-local collision terms needs to be scrutinized in detail which we leave for future studies.
\section{Summary and Conclusions}
\label{sec:summary}
In this work, we have extended the formalism presented in Ref.~\cite{Florkowski:2018fap} using the Wigner function formalism and employed semi-classical expansion to derive generalized quantum kinetic equations for the components of the Wigner function of massive spin-half Dirac particles. In our calculations, we have considered that the spin polarization effects may arise from both the zeroth and first-order contributions in the $\hbar$ expansion. 
We have derived a general quantum kinetic equation for the independent Wigner function components and used the ansatz for the generalized phase-space distribution function, including the spin degrees of freedom, to obtain its classical counterpart. We have shown that the latter has the same Boltzmann-like form as the one found in Refs.~\cite{Weickgenannt:2020aaf,Weickgenannt:2021cuo}, and the distribution function can be mapped back to the components of the Wigner function. In the present manuscript, we have not given explicit expressions of the collision kernel. However, we expect that using some suitable approximation of the collision kernel one can develop a general spin-hydrodynamic formalism in a way presented in Ref.~\cite{Weickgenannt:2022zxs} ensuring the conservation of the energy-momentum and total angular momentum tensors.
\section*{Acknowledgments}
We thank S.~Bhadury, A.~Jaiswal, D.~Rischke, E.~Speranza, and D.~Wagner for insightful discussions and N.~Weickgenannt for critical reading of the manuscript.
R.S., R.R., W.F., and A.D. were supported in part by the Polish National Science Centre Grants No.
2016/23/B/ST2/00717, No. 2018/30/E/ST2/00432, No. 2020/39/D/ST2/02054, and No. 2022/47/B/ ST2/01372. R.S. acknowledges the support of NAWA PROM Program No. PROM PPI/PRO/2019/1/00016/U/001 and the hospitality of the Institute for Theoretical Physics, Goethe University Germany where part of this work was completed, and Polish NAWA Bekker program No. BPN/BEK/2021/
1/00342. R.S. also thank the Institute for Nuclear Theory at the University of Washington for its kind hospitality and stimulating research environment. This research was supported in part by the INT's U.S. Department of Energy grant No. DE-FG02-00ER41132.
%
\begin{appendix}
\section{Third-order kinetic equations for the Wigner function coefficients}
\label{Asec:2}
In this appendix, we derive the third-order kinetic equations for the coefficients of the Wigner function. Mass-shell conditions and transport equations for $\mathcal{F}^{(2)}$ and $\mathcal{A}_{\mu}^{(2)}$ are also shown.
\smallskip

In the $\hbar^3$ order, comparing the real
and imaginary parts of the coefficients of the Wigner function in the Clifford-algebra basis we arrive at two sets of equations, where the real part gives,
\ba
k^{\mu}\mathcal{V}_{\mu}^{(3)}-m\mathcal{F}^{(3)}&=&\mathcal{D}_{\mathcal{F}}^{(2)},\label{equ68.1}\\
\frac{1}{2}\partial^{\mu}\mathcal{A}_{\mu}^{(2)}+m\mathcal{P}^{(3)}&=&-\mathcal{D}_{\mathcal{P}}^{(2)},\label{equ69.1}\\
k_{\mu}\mathcal{F}^{(3)}-\frac{1}{2}\partial^{\nu}\mathcal{S}_{\nu\mu}^{(2)}-m\mathcal{V}_{\mu}^{(3)}&=& \mathcal{D}_{\mathcal{V},\mu}^{(2)},\label{equ70.1}\\
-\frac{1}{2}\partial_{\mu}\mathcal{P}^{(2)}+k^{\beta}\hstar{\mathcal{S}}_{\mu\beta}^{(3)}+m\mathcal{A}_{\mu}^{(3)}&=&-\mathcal{D}_{\mathcal{A},\mu}^{(2)},\label{equ71.1}\\
\frac{1}{2}\partial_{[\mu}\mathcal{V}_{\nu]}^{(2)}-\epsilon_{\mu\nu\alpha\beta}k^{\alpha}\mathcal{A}^{\beta(3)}-m\mathcal{S}_{\mu\nu}^{(3)}&=&\mathcal{D}_{\mathcal{S},{\mu\nu}}^{(2)}\label{equ72.1}.
\ea
and from the imaginary parts we obtain
\ba
\partial^{\mu}\mathcal{V}_{\mu}^{(2)}&=&2 \mathcal{C}_{\mathcal{F}}^{(2)},\label{equ73.1}\\
k^{\mu}\mathcal{A}_{\mu}^{(3)}&=&\mathcal{C}_{\mathcal{P}}^{(2)},\label{equ74.1}\\
\frac{1}{2}\partial_{\mu}\mathcal{F}^{(2)}+k^{\nu}\mathcal{S}_{\nu\mu}^{(3)}&=&\mathcal{C}_{\mathcal{V},\mu}^{(2)},\label{equ75.1}\\
k_{\mu}\mathcal{P}^{(3)}+\frac{1}{2}\partial^{\beta}
\hstar{\mathcal{S}}_{\mu\beta}^{(2)}&=&-\mathcal{C}_{\mathcal{A},\mu}^{(2)}, \label{equ76.1}\\
k_{[\mu}\mathcal{V}_{\nu]}^{(3)} +\frac{1}{2}\epsilon_{\mu\nu\alpha\beta}\partial^{\alpha}\mathcal{A}^{\beta(2)}&=&- \mathcal{C}_{\mathcal{S},\mu\nu}^{(2)}.\label{equ77.1}
\ea
From Eqs.~\eqref{equ69}, \eqref{equ70} and \eqref{equ72} we obtain the second-order contributions to the pseudo-scalar, vector and tensor components of the Wigner function, respectively, 
\ba
\mathcal{P}^{(2)}&=&-\frac{1}{2m}\Bigg[\partial^{\mu}\mathcal{A}_{\mu}^{(1)}+2\mathcal{D}_{\mathcal{P}}^{(1)}\Bigg]\label{equ78},\\
\mathcal{V}_{\mu}^{(2)}&=&\frac{1}{m}\Bigg[k_{\mu}\mathcal{F}^{(2)}-\frac{1}{2}\partial^{\nu}\mathcal{S}_{\nu\mu}^{(1)}-\mathcal{D}^{(1)}_{\mathcal{V},\mu}\Bigg]\label{equ79},\\
\mathcal{S}_{\mu\nu}^{(2)}&=&\frac{1}{2m}\Bigg[\partial_{[\mu}\mathcal{V}_{\nu]}^{(1)}-2\epsilon_{\mu\nu\alpha\beta}k^{\alpha}\mathcal{A}^{\beta}_{(2)}-2\mathcal{D}_{\mathcal{S},{\mu\nu}}^{(1)}\Bigg].~~~~\label{equ80}
\ea
Again, the dual form of $\mathcal{S}_{\mu\nu}^{(2)}$ can be expressed as, 
\ba
\hstar{{\cal S}}_{\mu\beta}^{(2)}&=&\frac{1}{m}\left[\frac{1}{4}\epsilon_{\mu\beta \gamma \delta} \partial^{[\gamma}\mathcal{V}^{\delta]}_{(1)}
+ k_{[\mu}\mathcal{A}_{\beta]}^{(2)}
-\frac{1}{2}\epsilon_{\mu\beta \gamma \delta }\mathcal{D}_{\mathcal{S}(1)}^{\gamma\delta}\right].\nn\\
\label{equ81}
\ea
Contracting Eq.~\eqref{equ70} with $k^{\mu}$ and using Eqs.~\eqref{equ57} and \eqref{equ68} we get constraint equation for the second-order scalar coefficient, 
\ba
(k^2-m^2)\mathcal{F}^{(2)} &=& \frac{1}{4}\partial^{\mu}\partial_{\mu}\mathcal{F}^{(0)} + k^{\mu}\mathcal{D}_{\mathcal{V},\mu}^{(1)}+m\mathcal{D}_{\mathcal{F}}^{(1)}
-\frac{1}{2}\partial^{\nu}\mathcal{C}_{\mathcal{V},\nu}^{(0)}\,.
\label{equF2}
\ea
Using Eqs.~\eqref{equ60} and \eqref{equ81} in Eq.~\eqref{equ71}, we obtain the constraint equation for second-order axial-vector coefficient,
\ba
   &&(k^2-m^2)\mathcal{A}_{\mu}^{(2)}=\frac{1}{4}\partial_{\mu}\partial^{\alpha}\mathcal{A}_{\alpha}^{(0)}+ \frac{1}{2}\partial_{\mu}\mathcal{D}_{\mathcal{P}}^{(0)} +
   k_{\mu}\mathcal{C}_{\mathcal{P}}^{(1)}  \nn\\
   &&~~~~~~~+\frac{1}{2}\epsilon_{\mu\beta \gamma\delta} k^{\beta}\left(\partial^\gamma\mathcal{V}^\delta_{(1)}
   - \mathcal{D}_{\mathcal{S}(1)}^{\gamma\delta} \right) + m\mathcal{D}_{\mathcal{A},\mu}^{(1)}.\label{eq:Amu2}
\ea
Combining Eqs.~\eqref{equ73.1} and~\eqref{equ79} we get the kinetic equation for $\mathcal{F}^{(2)}$  as
\beq
   k^{\mu}\partial_{\mu}\mathcal{F}^{(2)}=2m\mathcal{C}_{\mathcal{F}}^{(2)}+\partial^{\mu}\mathcal{D}_{\mathcal{V},\mu}^{(1)}.\label{equ66.2}
\eeq
Finally, to arrive at the kinetic equation for  $\mathcal{A}_{\mu}^{(2)}$, we put Eqs.~\eqref{equ69.1} and \eqref{equ81} in Eq.~\eqref{equ76.1}, getting
\beq
k^{\beta}\partial_{\beta}\mathcal{A}_{\mu}^{(2)}=2m\mathcal{C}_{\mathcal{A},\mu}^{(2)}-2k_{\mu}\mathcal{D}_{\mathcal{P}}^{(2)}-\frac{1}{2}\epsilon_{\mu\beta \gamma\delta}\partial^{\beta}\mathcal{D}_{\mathcal{S}(1)}^{\gamma\delta}.
~~~~~~\label{equ87}
\eeq
\end{appendix}
\bibliography{bibliography}{}
\bibliographystyle{utphys}
\end{document}